%% file: Asilomar2021.tex
\documentclass[conference]{IEEEtran}
\IEEEoverridecommandlockouts
\usepackage{cite}
\usepackage[numbers]{natbib}
\usepackage{amsmath,amssymb,amsfonts}
\usepackage{algorithmic}
\usepackage{graphicx}
\usepackage{textcomp}
\usepackage{xcolor}
\usepackage{siunitx}
\DeclareSIUnit{\decibelm}{dBm}
\DeclareSIUnit{\decibeli}{dBi}
\DeclareSIUnit{\farad}{F}
\usepackage{booktabs}
\usepackage[]{authblk}

\usepackage{glossaries}
\loadglsentries{abbreviations}

\def\BibTeX{{\rm B\kern-.05em{\sc i\kern-.025em b}\kern-.08em
    T\kern-.1667em\lower.7ex\hbox{E}\kern-.125emX}}
    
\makeatletter
\def\blfootnote{\gdef\@thefnmark{}\@footnotetext}
\makeatother

\usepackage[pscoord]{eso-pic}
\newcommand{\placetextbox}[3]{
  \setbox0=\hbox{#3}
  \AddToShipoutPictureFG*{
    \put(\LenToUnit{#1\paperwidth},\LenToUnit{#2\paperheight}){\vtop{{\null}\makebox[0pt][c]{#3}}}%
  }%
}%

\begin{document}

\title{A Multi-band Solution for Interacting with Energy-Neutral Devices

\thanks{The project has received funding from the European Union’s Horizon 2020 research and innovation programme under grant agreement No 101013425.}
}

\author[]{Chesney Buyle}
\author[]{Bert Cox}
\author[]{Liesbet Van der Perre}
\author[]{Lieven De Strycker}
\affil[]{Department of Electrical Engineering, KU Leuven, Belgium}


\maketitle


\begin{abstract}
RF Wireless Power Transfer (WPT) emerges as a technology for charging autonomous devices, enabling simultaneous power and information transfer. However, with increasing distance, single-input, single-channel rectenna systems are not able to meet the power requirements of large scale IoT applications. In this paper, we tackle this problem on two levels. First, we minimize the energy consumption at the energy-constrained device on three levels. Second, we evolve to a dual-band solution increasing RF WPT. One frequency band is used to provide a base charge to many nodes in a shared transmission. Beam steering, on the other hand, allows for more power hungry operations while introducing as minimal interference as possible. We showcase this method for a hybrid RF-acoustic positioning system. Practical measurements conducted in a multi-antenna indoor testbed (Techtile) show the additional power gain and positioning rate.  
\end{abstract}



\section{Introduction}



\placetextbox{0.5}{0.07}{
    \begin{footnotesize}
        This work has been submitted to the IEEE for possible publication. Copyright may be transferred without notice, after which this version may no longer be accessible.
    \end{footnotesize}
}

With the Internet of Things (IoT) becoming a well adopted term in mainstream business and the prospection of 43 billion connected devices in 2023~\cite{McKinsey2019}, sensor technology partially puts pressure on reliable and sustainable energy. Acknowledging the Sustainable Development Goals~\cite{SDG_2015}, higher maintenance and device costs, academic research should take the lead in dissuading battery-driven sensor solutions and invest time in exploring large scale, energy-neutral resolutions. Two main research paths can be distinguished~\cite{Ejaz2017}: optimizing energy consumption on the devices by means of improved hardware design~\cite{Aras2019, Thoen2017}, communication schemes~\cite{Wielandt2020} or data compression~\cite{Stojkoska2017} and ambient energy harvesting, where power is scavenged from the surroundings. This paper will focus on the latter. For outdoor applications, efficient energy harvesting can be obtained through small scale renewable energy sources such as solar panels and wind turbines. These sources are commonly chosen for their high power density~\cite{Sangkil2014}. In indoor environments, wireless sensor networks do not have access to these energy sources and have to rely on less efficient methods for battery free operation. A solution with great potential can be found in radio frequency (RF) harvesting~\cite{Lu2015} as it can deliver energy as well as transport information simultaneously (SWIPT)~\cite{Varshney2008}. The lower efficiency and the reciprocal relation to the distance of this far-field method are countered by the ever expanding wireless communication infrastructure, increasing the potential energy sources.

In previous research~\cite{Cox2020}, we showed that directional, single band wireless power transfer in combination with RF backscattering enables the full passive positioning of a hybrid RF-acoustic mobile tag. However, two problems raised, namely the low update rate and the position presumption for directional power transfer. This latter is a classical `chicken or egg' causality dilemma: to steer a beam towards the mobile tag, a position has to be known, but to calculate a position, the mobile tag needs to have power. The present study tries to solve this dilemma by implementing a dual band solution, where an omnidirectional antenna will provide energy to all mobile tags, providing enough energy to calculate its position every once in a while. A secondary directional antenna can be focused on certain tags to boost their energy budget and subsequently increase the positioning update rate

Extensive research on dedicated antenna, matching network, rectifier and architecture design have been described in the overview paper of~\cite{Lu2015}. Together with recent developments on dual band~\cite{LiLong2021, LiShun2021},  multi-\cite{Vu2020} and broadband~\cite{Liu2020} systems, it is shown that a high efficiency (up to 71\%) can be obtained with these dedicated designs. In this paper we tackle dual band RF energy harvesting by a more practical implementation approach. An off-the-shelf energy harvesting chipset~\cite{AEM4940}, containing the necessary rectifier, converters and energy storage controllers, is used in combination with commercially available dipole antennas. An energy harvesting performance analysis is performed in two bands, namely the \SI{868}{\mega\hertz} and \SI{2.4}{\giga\hertz} RFID bands and follow the specific restrictions regarding the ETSI regulations in each band. As an energy neutral devices (END), in house developed audio backscatter, indoor positioning hardware is used. Based on key parameters affecting the energy consumption of the END, we calculate the necessary energy budget and evaluate the position update rate in function of the distance through theoretical calculations and practical measurements.

The next section discusses the in-house developed hybrid RF-acoustic ranging hardware. Section III focuses on the dual band charging setup and potential energy harvesting performance in the \SI{868}{\mega\hertz} and \SI{2.4}{\giga\hertz} RFID bands. A theoretical and practical performance assessment is made in sections IV and V respectively. Conclusions and future work are presented in section VI. 

\section{Energy-neutral operation: energy budget analysis}
\subsection{Ultra-Low Energy Audio Backscatter Hardware}
One of the prerequisites for the intended directional RF energy transfer is that the position of the mobile tag needs to be known, enabling the system to steer the beam into the direction of the tag. Additionally, the energy consumption of this tag should be as low as possible for a battery free operation. These conditions are fulfilled by in-house developed hybrid RF-acoustic hardware. Distance measurements, and consequently indoor positioning is done by awaking the mobile tags for a short time ($\tau_{rx}$ = 1\,ms), and communicating the received part of an ultrasonic chirp at that time back to the central computing system. Depending on the distance to the audio source, another part of the chirp is received at the tag, and through cross correlation, this distance can be estimated. Communication is done over RF using backscattering, where an incoming RF wave is reflected or absorbed according to the acoustic data. \newline
From the mobile tag's perspective, three key parameters can be defined that affect the energy consumption:
\begin{itemize}
    \item \textbf{The hardware power consumption ($P$)}. The most obvious parameter, nevertheless, a great deal of hardware design effort should be invested in this as a large energy gain can be achieved by a simple job like selecting the right component.
    \item \textbf{Audio reception window ($\tau_{rx})$}. The main advantage of increasing this window is the improved ranging accuracy. Pulse compression through cross correlation results in better defined maxima and consequently more precise distances. However, during this wake-on period, the different hardware blocks will consume the aforementioned power ($P$). In the current set-up, the reception window is fixed to 1\,ms.
    \item \textbf{The turn-on time ($\Delta t_{ON}$)}. A common practice within low-power electronics is to switch to a deep sleep state when not being used. Power consumption in this state is low, but can not be neglected, certainly in cases where the sleep time is substantially larger than the awake time. For this reason, the power is cut off outside the audio reception window. The turn-on time is defined as the time it takes before there is any output when the power is switched on. Again, great care should be taken into hardware design as this wake-up time is component specific and large current spikes can be observed in this stage. 
\end{itemize}
Power and timing measurements show that a single ranging estimation costs 3.15\,µJ at the mobile tag side (= $E_{tag}$). This directly defines the harvesters' necessary energy storage size: $E_{store} > 4 \cdot E_{tag}$ (four distance measurements results in one 3D position estimation).

\subsection{Harvesting chipset}
\begin{figure}[]
 \centering
\includegraphics[width=0.45\textwidth, angle=0]{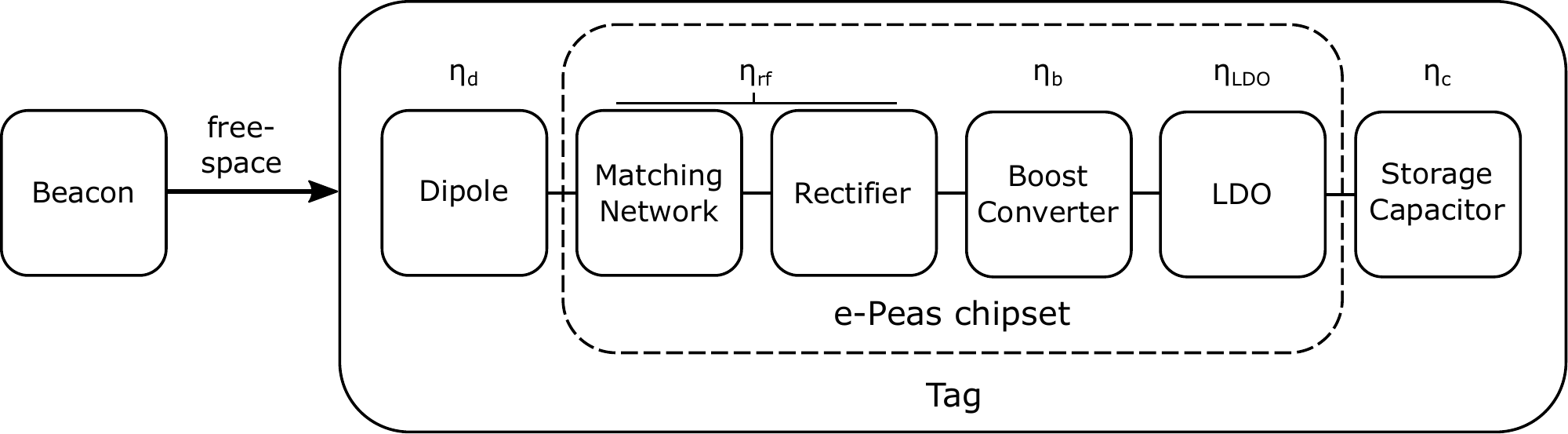}
\caption{Overview of the link budget elements and their efficiencies.}
\label{fig:simulation}

\end{figure}

The commercial integrated energy management chipset used in this assessment is the E-peas AEM40940~\cite{AEM4940}. According to the datasheet, it extracts AC power from high-frequency RF inputs to simultaneously store energy in a rechargeable element and supply the system with two independent regulated voltages. An overview of its building blocks, the link budget and corresponding efficiencies is given in Fig.~\ref{fig:simulation}. A capacitor is chosen as energy storage element due to the limited energy budget at the mobile tag and its fast charging capabilities. An RF input power between -19\,dBm and 10\,dBm is necessary at the input of the energy harvester to start charging the capacitor. The size of the capacitor depends on two voltage levels:  $V_{chrdy}$ is the minimum starting voltage on which the energy stored in the capacitor can be used. The voltage over this capacitor slowly degrades until it reaches $V_{ovdis}$, at this point it stops delivering power to the positioning hardware. These two voltages are the boundaries on which power will be delivered and are subject to the hybrid RF-acoustic hardware. The capacitance is set to \SI{22}{\micro\farad} according to~\cite{Cox2020} with $V_{chrdy}$ = \SI{3.10}{\volt}, $V_{ovdis}$ = \SI{2.8}{\volt} and worst case LDO efficiency $\eta_{LDO}$ of \SI{74}{\percent}. The necessary storage energy to be harvested can be calculated with:
\begin{equation}
    E_{store}  = \frac{C \cdot (V_{chrdy}^2 - V_{ovdis}^2)}{2} = \SI{19.5}{\micro\joule} > 4 E_{tag}
\end{equation}

Table~\ref{table:efficiencies} summarizes the different efficiencies of the link budget in the receiver. Both the efficiency of the dipole ($\eta_d$) and the internal RF path efficiency ($\eta_{rf}$) are frequency dependent. The first one defines how well the supplied power is converted into radiated power, and depends on conduction and dielectric losses. The efficiency of an antenna can be read directly from the datasheets for the used 868\,MHz~\cite{TI852113} and 2.4\,GHz~\cite{GW11A153} antenna and is slightly higher for the 2.4\,GHz. The RF-path efficiency is defined as the ratio between the power coming out of the rectifier and the power entering the RF-matching network. Even with a tuned RF matching network, it can be noted that the efficiency is lower at the higher frequency. The internal boost converter ($\eta_b$) and LDO ($\eta_{LDO}$) lower the total efficiency at the same rate. Lastly, there are almost no losses in the charging capacitor. This is due to the low Equivalent Series Resistor (ESR) of the chosen aluminum polymer capacitor, which has the largest influence on the efficiency of a capacitor as storage element.
\begin{table}[!htb]
\caption{Efficiencies in the receiver's link budget for a receive power of -10\,dBm.}
\label{table:efficiencies}
\renewcommand{\arraystretch}{1.5}
\centering
\begin{tabular}{lccccc} 
\toprule
\textbf{Frequency} & $\eta_{d}$ (\%) & $\eta_{rf}$ (\%) & $\eta_{b}$ (\%) & $\eta_{LDO}$ (\%) & $\eta_{c} (\%) $\\ \hline
868\,MHz & 75.17 & 37.78 & 73.80 & 74.07 & 99  \\
2.4\,GHz & 80.00 & 14.13 & 73.80 & 74.07 & 99 \\ 
\bottomrule
\end{tabular}
\end{table}

\section{System overview and regulation analysis}

\subsection{Dual band charging setup}

\begin{figure}[]
 \centering
\includegraphics[width=\linewidth, angle=0]{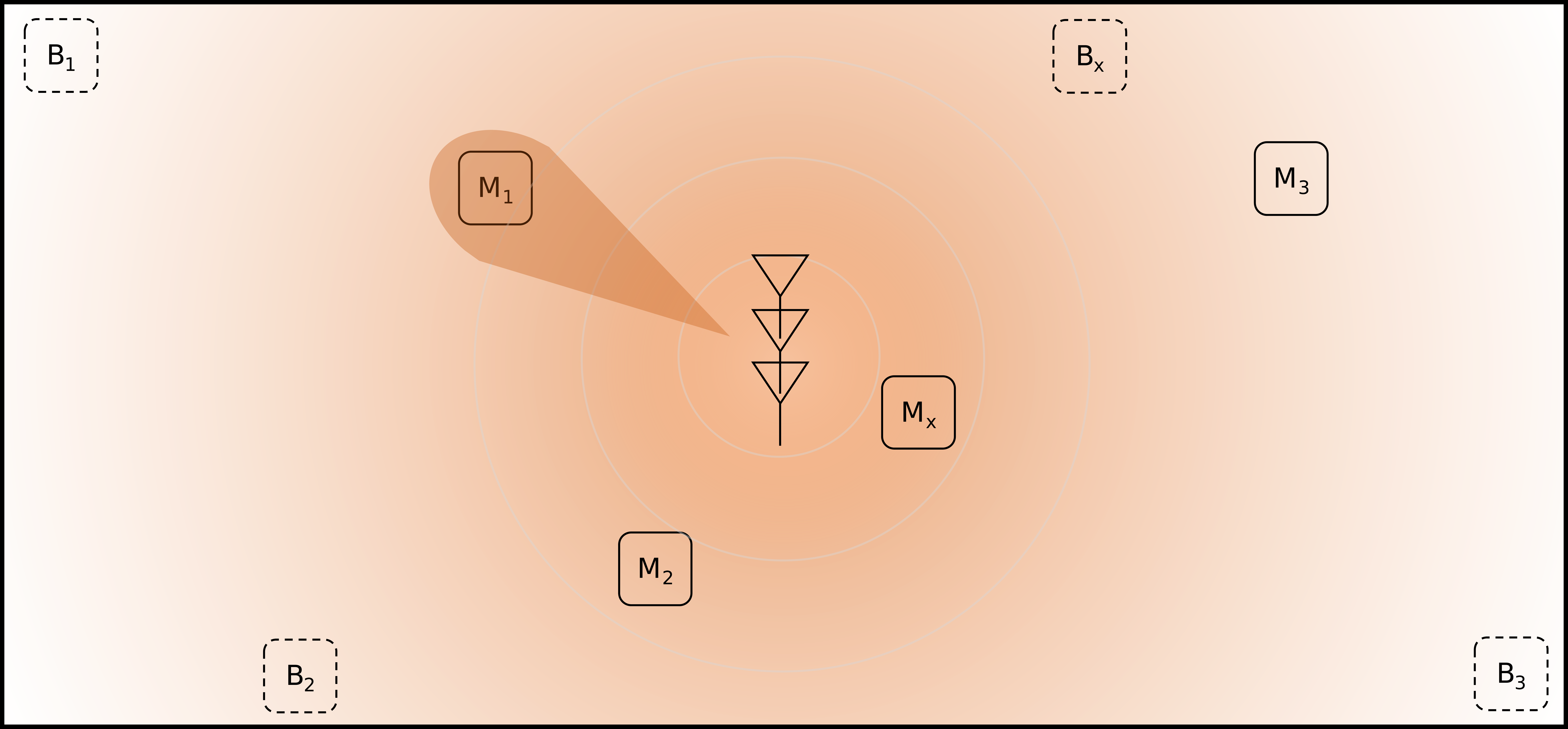}
\vspace{-0.2cm}
\caption{The dual band, multi-antenna charging setup}
\label{fig:setup}
\end{figure}

A potential setup of the RF power transfer system is shown in Fig.~\ref{fig:setup}. It consists of the following three entities: an RF power transmitter, one or multiple mobile tags (M) and several speaker beacons (B). The power transmitter is composed of two separate antenna systems. On the one hand, a single omnidirectional antenna is used to provide a base charge to the mobile tags. In this way, a guaranteed charge is delivered to all the tags in the vicinity to power up once in a while. The antenna is located at the center of the room to minimize the distance and consequently path loss to the surrounding tags. On the other hand, an antenna array or rotating directional antenna is introduced to steer a high power beam towards one or more specific tags. Directional power transfer offers two main advantages over omnidirectional transmission: (I) a higher amount of power can be transferred to the mobile tags. Moreover, the energy harvester's efficiency increases overall with increasing input power. (II) Interference with communications systems is reduced since less multipath reflections are present. Once again, the directional steering system is positioned at the center of the room to limit the TX-RX distance. 

In order to enable simultaneous operation of the omnidirectional and directional power transfer system, they must operate in different frequency bands. The E-peas AEM40940 supports energy harvesting in two European ISM bands, namely \SI{868}{\mega\hertz} and \SI{2.4}{\giga\hertz}. The design of the dual band charging setup will mainly depend on two parameters: the energy conversion efficiency and (inter)national RF transmission regulations. For example, a high energy conversion efficiency and transmission power are desired for both the omnidirectional and directional power transfer systems as a higher amount of energy can be delivered to the mobiles tags. On the other hand, the requirements regarding the RF transmission duty cycle may be more asymmetric for the two frequency bands: the omnidirectional antenna system is deployed to provide a continuous base charge to all the mobile tags in the vicinity, while the directional boost charge must only be available when a higher position update rate is required. In the worst case scenario, the directional power transfer system also requires continuous transmission. 

\subsection{Regulation analysis}
The frequency bands considered for RF power transfer are the \SI{868}{\mega\hertz} and \SI{2.45}{\giga\hertz} RFID bands as they allow transmissions with relatively high transmit power and duty cycle. The harmonised European standards for transmission in the \SI{868}{\mega\hertz} and \SI{2.45}{\giga\hertz} RFID bands can be found in respectively ETSI EN 302 208 V3.1.1 and ETSI EN 300 440 V2.2.1. 

Transmission in the \SIrange[range-phrase=-]{865}{868}{\mega\hertz} lower RFID band can be done in any of the four available high power channels. The maximum \gls{erp} depends on the beamwidth of the transmit antenna in the horizontal orientation. In the case of omnidirectional transmission, the \gls{erp} shall be \SI{500}{\milli\watt} at maximum. When the beamwidth is less than or equal to \SI{180}{\degree} or \SI{90}{\degree}, the \gls{erp} of the transmitter can be increased to a maximum of \SI{1}{\watt} and \SI{2}{\watt} respectively. In the preferred method of operation, the length of transmission should be kept to a minimum and should be as long as is required to read the tags in the environment and check whether no additional tags are present. However, continuous transmission is preferred for power transfer to keep the mobile tags charged at all times. Therefore, transmitters may operate in a presence-sensing mode in which they perform periodic transmissions to verify if new mobile tags are present. In this mode, continuous transmissions of max. \SI{1}{\second} are interleaved with silent periods of at least \SI{100}{\milli\second}. 

For transmission in the \SI{2.45}{\giga\hertz} RFID band, two main transmission limits are defined. If the \gls{eirp} of the transmitter is limited to \SI{27}{\decibelm}, unmodulated carrier (CW) or \gls{fhss} signals can be transmitted without restrictions on duty cycle. The \gls{eirp} may be increased up to \SI{36}{\decibelm} if the equipment is used in in-building environments and  \gls{fhss} techniques are employed. However, the average duty cycle of the transmission is limited to \SI{15}{\percent} over any period of \SI{200}{\milli\second}. Moreover, in both cases, the antenna must have a horizontal beamwidth of \SI{45}{\degree}, a sidelobe attenuation of at least \SI{15}{\decibel} and physical protections with dimension limits. 

Aforementioned transmission regulations show that transmission in the \SIrange[range-phrase=-]{865}{868}{\mega\hertz} RFID band is preferred for both the omnidirectional and directional power transfer. On the one hand, the antenna beamwidth restriction in the \SI{2.45}{\giga\hertz} RFID band prohibits omnidirectional charging of the mobile devices. On the other hand, although the maximum transmission power is lower in the \SIrange[range-phrase=-]{865}{868}{\mega\hertz} RFID band, the higher duty cycle limit ensures a higher energy transfer to the mobile tags overall. Moreover, the higher RF path efficiency ($\eta_{rf}$) of the E-peas energy harvester at the frequency of \SI{868}{\mega\hertz} compared to the \SI{2.4}{\giga\hertz} frequency band ensures an increased power transfer.

\begin{figure}[]
 \centering
\includegraphics[width=0.9\linewidth, angle=0]{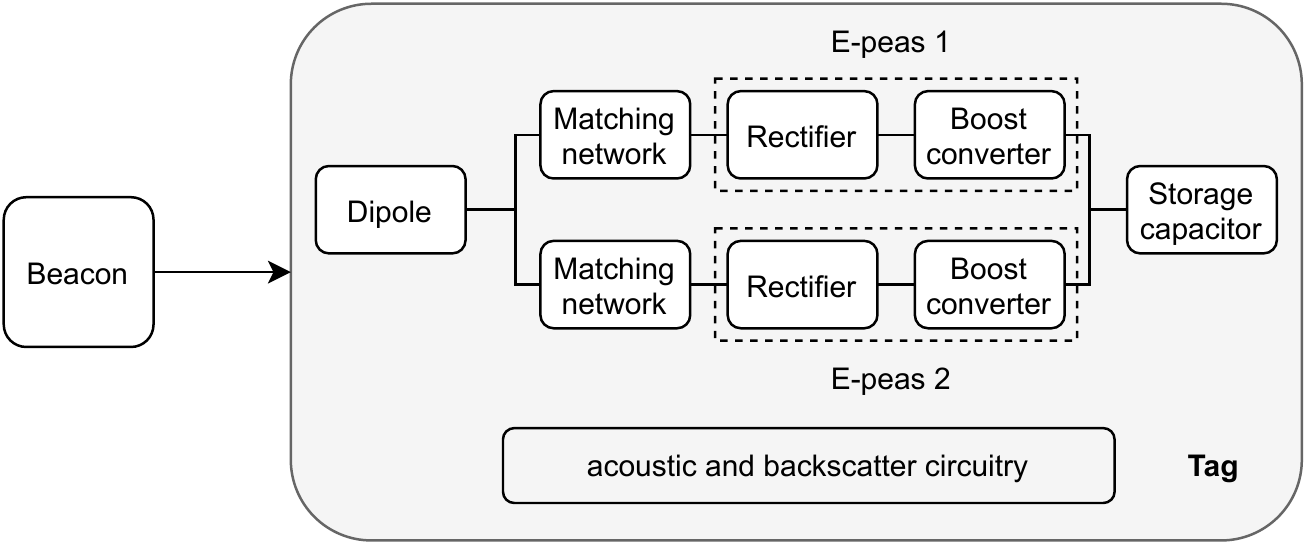}
\caption{Proposed dual band energy harvesting system of the tag.}
\label{fig:dualbandEHsystem}
\end{figure}

In conclusion, the presented RF WPT system operates completely within the \SIrange[range-phrase=-]{865}{868}{\mega\hertz} RFID band. However, simultaneous operation of the omnidirectional base charging and directional boost charging subsystems is enabled by facilitating simultaneous transmission and energy harvesting in the different high power sub-bands. The proposed internal structure of the tag's energy harvesting circuitry is shown in Fig.~\ref{fig:dualbandEHsystem}. Two parallel energy harvester stages are connected to a single dipole antenna. In order to ensure an even distribution of the spectrum, the matching network of each stage could be tuned to half of the four available high power channels. Both energy harvesting stages are connected to the same storage capacitor of which the acoustic and backscatter circuitry is powered. 

\section{Theoretical performance assessment}

\begin{figure}[]
 \centering
\includegraphics[width=0.95\linewidth, angle=0]{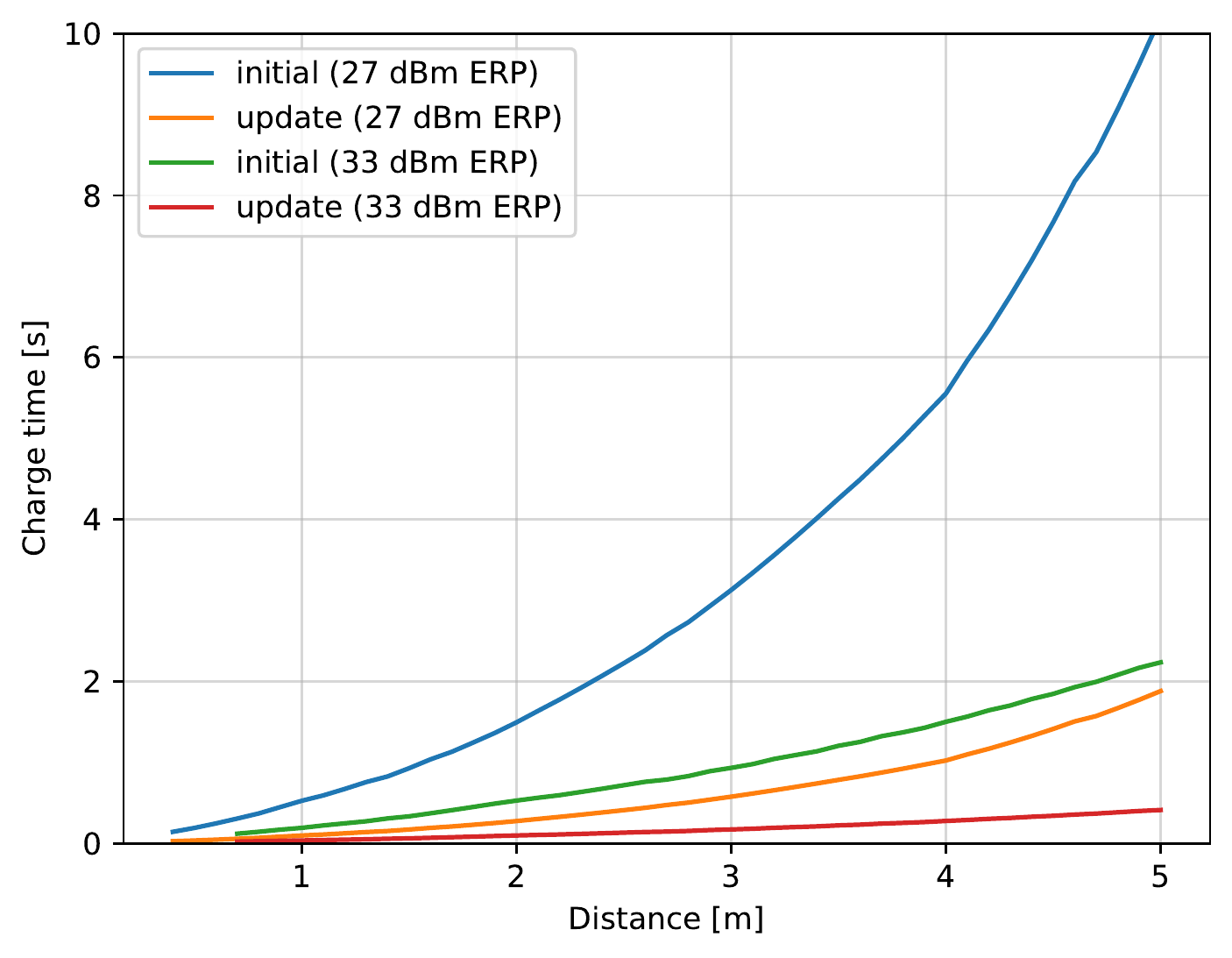}
\vspace{-0.2cm}
\caption{Simulated initial and update charge time in function of the distance.}
\label{fig:sim_charge_time}
\end{figure}

A performance analysis of the RF WPT system is made in terms of the tag's charge time, i.e. the time it takes for the transmit beacon to charge the tag's storage capacitor with sufficient energy to perform four ranging measurements, or one 3D position estimation. We consider two performance parameters: 
\begin{enumerate}
	\item Initial charge time = time required for the energy harvester to charge the storage capacitor from \SI{0}{V} to $V_{chrdy} = \SI{2.3}{V}$. This scenario takes place when the tag has not been charged for a long time. 
	\item Update charge time = time required for the energy harvester to charge the storage capacitor from $V_{ovdis} = \SI{2.8}{V}$ to $V_{chrdy} = \SI{3.1}{\volt}$. This scenario takes place when the tag was recently or is continuously powered. 
\end{enumerate}

The calculations of these charge times are based on the Friis transmission formula and E-peas AEM40940 energy harvesting efficiency. At the transmit beacon, we consider a continuous wave (CW) at the frequency of \SI{865.7}{\mega\hertz}. The \gls{erp} is either limited to \SI{27}{\decibelm} or \SI{33}{\decibelm}, depending on the transmit antenna's beamwidth. No transmit duty cycle limitations are considered. At the tag, we assume a dipole antenna with \SI{2.15}{\decibeli} gain. In accordance with Fig.~\ref{fig:simulation}, the matching network ($\eta_m$), rectifier ($\eta_r$), boost converter ($\eta_b$) and LDO efficiency ($\eta_{LDO}$) are provided by the E-peas AEM40940 technical datasheet~\cite{AEM4940}. No losses are assumed in the charging process of the storage capacitor ($\eta_c = 1$). 

Fig.~\ref{fig:sim_charge_time} shows the simulated initial and update charge times in function of the beacon-tag distance for the two considered \gls{erp} scenarios. When a mobile tag enters the RF field of the WPT system with a depleted storage capacitor, it would take the omnidirectional base charging system around \SI{10}{\second} to power up the tag at a distance of \SI{5}{\meter}. Once the initial charge is overcome, the base charging system is able to initiate an updated 3D position measurement every \SI{2}{\second} for the same distance. If a higher update rate is desired, the directional boost charging system can be used. In this case, it takes only around \SI{0.4}{\second} to obtain an updated 3D position from the tag at \SI{5}{\meter} distance.

\section{Practical performance assessment}
\subsection{Measurement Setup}

A practical performance evaluation of the RF WPT system was performed in the Techtile environment, a wooden construction of 8\,x\,4\,x\,2.4\,m~\cite{callebaut2021primer}. A dipole~\cite{TI852113} and directional patch (\SI{5}{\decibeli} gain) antenna were alternately placed in the center of the room at a height of \SI{1}{\meter} as transmit beacon. A sinuso\"idal continuous wave of frequency \SI{865.7}{\mega\hertz} was transmitted by the beacon. At the mobile device, the same dipole antenna was connected to an E-peas AEM40940 energy harvester board, which was loaded with our custom hybrid RF-acoustic ranging module. Opposed to the proposed dual band energy harvesting system presented in Fig.~\ref{fig:dualbandEHsystem}, a single E-peas energy harvester was used to evaluate the omnidirectional and directional WPT measurements separately. Preliminary tests showed that a slightly larger aluminum electrolytic capacitor of \SI{39.7}{\micro\farad} had to be used to provide sufficient energy to the RF-acoustic module in order to perform the four ranging measurements, and thus single 3D position estimation. 

Four measurement campaigns were conducted. The first two consisted of measuring the initial charge time for both the omnidirectional and directional antenna configuration. This is the time interval measured between the start of RF transmission at the beacon and the moment at which the energy harvester charged its storage capacitor from \SI{0}{\volt} to $V_{chrdy}$. The third and fourth campaign evaluated the update charge time of the WPT system, i.e. the time interval measured between the start of RF transmission at the beacon and the moment at which the energy harvester charged its storage capacitor from $V_{ovdis}$ to $V_{chrdy}$. In accordance with the ETSI regulations, the \gls{erp} for the omnidirectional antenna configuration was set to \SI{27}{\decibelm}. In the case of the directional patch antenna configuration, the \gls{erp} was set to only \SI{31.9}{\decibelm} due to limitations of the used RF power amplifier. In each scenario, the distance between the transmitter and receiver was sequentially increased in steps of \SI{25}{\centi\meter}, starting from \SI{0.5}{\meter} and up to the border of the Techtile room at \SI{4}{\meter}. At each distance, the considered charging time is measured 25 times. The measurements using the omnidirectional and directional antenna configurations were conducted separately in order to prevent interference. 


\subsection{Results}

\begin{figure}[]
 \centering
\includegraphics[width=0.95\linewidth, angle=0]{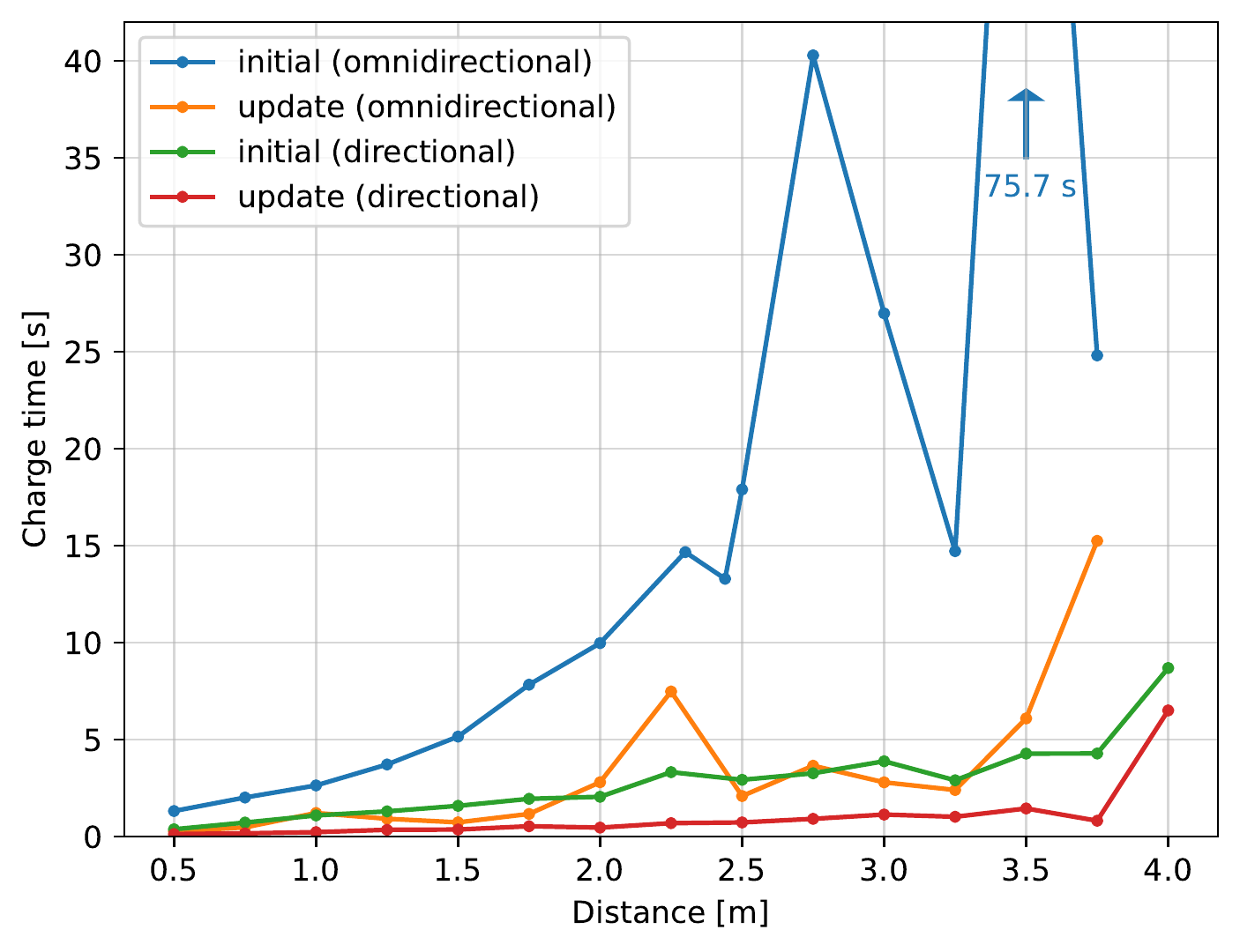}
\vspace{-0.2cm}
\caption{Measured initial and update charge time in function of the distance.}
\label{fig:measured_charge_times}
\end{figure}

Fig.~\ref{fig:measured_charge_times} shows the mean initial and update charge time of both transmit antenna configurations in function of the distance. The initial charge time lies in both cases substantially higher than their update charge time counterpart. In the most unfavorable scenario, omnidirectional transmission and initial charging, a position update can be achieved every \SI{40}{\second}, with exception at \SI{3.5}{\meter} distance.  Our initial guess is that multipath effects cause destructive interference but further investigations and measurements should test this hypothesis. The obtained position can then be used to steer the directional power beam towards the mobile device. In this case, the update charge time falls overall below \SI{2}{\second}, except at the distance of \SI{4}{\meter}.

\section{Conclusion}
In this research, we show that a 3D positioning update time of \SI{1}{\second} can be achieved with commercially available energy harvesting chipsets and within the ETSI regulations. A theoretical comparison between the potential energy harvesting performance in two ISM bands, namely the \SI{868}{\mega\hertz} and \SI{2.4}{\giga\hertz} RFID bands is made. It shows that the lower frequency band is more useful for energy harvesting because of two reasons: a higher RF path efficiency at the considered energy harvester chipset and lower duty cycle limitations. Practical measurements in a non-anechoic chamber show that when beam steering is introduced, the charge time can be up to 10 times lower. In this case, the energy storage element can be charged below \SI{10}{\second} at the edges of the testbed. In future investigations we will check if a custom, high frequency energy harvester can bridge the current gap, resulting in potentially smaller and more efficient hardware.


{\footnotesize
\bibliography{library.bib}}%
\bibliographystyle{IEEEtranN}%

\end{document}

%% file: abbreviations.tex
\newacronym{erp}{ERP}{Effective Radiated Power}
\newacronym{eirp}{EIRP}{Effective Isotropic Radiated Power}
\newacronym{fhss}{FHSS}{Frequency Hopping Spread Spectrum}